\documentclass[runningheads]{llncs}
\usepackage[T1]{fontenc}
\usepackage{graphicx}
\usepackage{arabtex}
\usepackage{hyperref}
\usepackage{amsmath}

\usepackage{color}
\usepackage{hyperref}
\hypersetup{
    colorlinks=true,
    linkcolor=blue,
    filecolor=magenta,      
    urlcolor=cyan,
    citecolor=blue,
    breaklinks=true,
    pdftitle={An Analysis of Prompting for Visual Art},
    pdfpagemode=FullScreen,
}

\usepackage{multirow}
\usepackage{booktabs}
\usepackage{xurl}
\usepackage{tabularx}
\graphicspath{{Figures/}}

\begin{document}
\title{No Longer Trending on Artstation: Prompt Analysis of Generative AI Art}
%
%
\author{Jon McCormack\orcidID{0000-0001-6328-5064} \and
Maria Teresa Llano\orcidID{0000-0002-4898-1755} \and
Stephen James Krol\orcidID{0000-0002-9474-3838} \and
Nina Rajcic\orcidID{0000-0001-6501-5754}}

\authorrunning{J. McCormack et al.}
%
%
\institute{SensiLab, Monash University, Caulfield East, Victoria 3145, Australia 
\email{\{Jon.McCormack, Teresa.Llano, Stephen.Krol, Nina.Rajcic\}@monash.edu}}

\maketitle              
\begin{abstract}
Image generation using generative AI is rapidly becoming a major new source of visual media, with billions of AI generated images created using diffusion models such as Stable Diffusion and Midjourney over the last few years. In this paper we collect and analyse over 3 million prompts and the images they generate. Using natural language processing, topic analysis and visualisation methods we aim to understand collectively how people are using text prompts, the impact of these systems on artists, and more broadly on the visual cultures they promote. Our study shows that prompting focuses largely on surface aesthetics, reinforcing cultural norms, popular conventional representations and imagery. We also find that many users focus on popular topics (such as making colouring books, fantasy art, or Christmas cards), suggesting that the dominant use for the systems analysed is recreational rather than artistic.  

\end{abstract}

\section{Introduction}
\label{s:introduction}

In just a few years, generative AI has become a dominant new paradigm for the creation of high-quality digital media, including images, video and text. One recent estimate put the number of AI generated images created to date at over 15 billion -- exceeding the number of photographs taken in the first 150 years of photography \cite{EveryPixel2023}. Text-to-image (TTI) systems, such as Stable Diffusion, Midjourney or DALL-E allow the creation of high-quality images or illustrations just by supplying a short text description (prompt) of the desired image.
These systems have become so popular, that ``prompting'' is now being considered as a part of the skillset of successful commercial art production.

However, as is the case with most recent generative AI advances, these systems raise many cultural, ethical and conceptual issues. Exploring and understanding these issues is increasingly important, given the rapid update and normalisation of generative AI imagery into mainstream software tools\footnote{Adobe's generative AI system ``firefly'' is now integrated into Photoshop, with users already generating over 2 billion images \cite{Adobe2023}.}. Important concerns raised by generative AI include bias and the reinforcement of cultural stereotypes, concerns regarding censorship, issues of data laundering \cite{Baio2022} and training on copyrighted data, ``style theft'' and the automation of specific artistic styles, questions of authenticity, ``hallucinations'' and ``deep-fake'' imagery, loss of traditional human skills, cultural and aesthetic homogenisation \cite{Crawford2021,McCormack:2023aa,Turk2023}. At the centre of many of these issues are the prompts themselves, the language they cultivate, and the \emph{type} of images they produce. By understanding how -- collectively -- people are expressing their ideas through prompts, we can improve our understanding of how generative AI is impacting visual art and culture.

In this paper we examine several large prompting datasets, using Natural Language Processing (NLP) and data visualisation to draw a broad understanding of how users of TTI systems utilise language in generative AI systems. Our overall aim is to improve understanding of generative AI's influence on how we think when making a prompt and in particular, our use of language in relation to art-making and creativity. Hence, we focus on how artistic concepts are represented in prompt language; tracing their evolution as TTI systems continue to become more prolific and technically sophisticated. To this end, we compare prompting datasets from mid-2022\footnote{In mid-2022, many of the current systems were in pre-release, or just gaining a sizeable body of users.} to November 2023, by which point, TTI systems had become widely adopted. We also explore how TTI systems have and continue to impact human artists and illustrators. 

\section{Background and Related Work}
\label{s:background}

\subsection{Text-to-Image Generation}

The introduction of diffusion models marked a significant leap in image generation quality over previous methods such as Generative Adversarial Networks (GANs) \cite{goodfellow2020generative}. The fundamental innovation of recent prompt-based image generators is the fusion of two independently powerful models -- CLIP  (Contrastive Language–Image Pre-training) \cite{radford2021learning} for text-based understanding, and Diffusion Models for image synthesis -- into a cohesive system capable of generating images from textual prompts. 

CLIP was trained on dataset of 400 million images and their associated textual descriptions. Unlike predecessor models that focused on direct object recognition, CLIP also set out to understanding the context, style, and other abstract concepts conveyed through language. The training set was scraped from online sources: social media posts, image-caption pairs, alt text descriptions, among others. For TTI models, CLIP serves as a mechanism to interpreting prompts. When a user inputs a textual description, CLIP analyses this text to construct the visual elements and themes that should be present in the output image.

\subsection{Prompt and Image analysis}
With the introduction of TTI systems, in addition to being a creative tool, generative AI has become a social phenomenon  \cite{sanchez:2023}. 
Writing prompts is primarily an iterative, trial and error process; an exploratory practice \cite{xie:2023}. More importantly, specific terms that coax the model into generating high-quality results have become a significant aspect of successful prompting.

Existing literature in prompt analysis has focused on aiding prompt engineering and making improvements to the underlying models \cite{xie2023}; for instance, by identifying popular or ``most useful'' keywords and model parameters that would allow users to generate higher quality images \cite{liu:2022,zamfirescu-pereira:chi23,pavlichenko:sigir23}. Xie at al \cite{xie:2023} further propose using higher rating prompts -- in which their analysis pointed at longer prompts, and prompts including artists' names -- as a way to further train the models, feeding back to the systems what users want. These approaches to prompt analysis  pay less attention to large-scale analysis from a cultural perspective, and often favour feedback approaches that tend to homogenise content and reinforce popular clichés, amplifying concerns about biases, stereotypes, authorship and authenticity. Additionally, they reinforce the artistic styles of just a handful of specific artists (see our analysis below). 

A recent study undertook a more exploratory approach to prompt analysis \cite{sanchez:2023} using the DiffusionDB database \cite{wang-etal-2023-diffusiondb}. Applying topic analysis, the author identified a taxonomy of prompt specifiers and developed a model for identifying the categories from this taxonomy from TTI prompts. Although the main goal of this study was the development of an interactive tool to assist with prompting, the results from this work open up possibilities for better understanding of what topics are being expressed in prompts. We build upon this work, performing topic analysis on a more recent dataset, and examining how current trends in prompting may inform thinking about art and visual culture. 

Visual analysis of AI-generated images has also become a topic of significant interest, with 
current studies focused on identifying biases (e.g. under-representing certain race groups \cite{bansal:emnlp2022,naik:aies23}), cultural gaps (e.g. over-representing specific nations \cite{naik:aies23}), and the reinforcement of stereotypes (e.g. ``a photo of a lawyer'' consistently showing a white male) \cite{bianchi:facct23}). An analysis of 3,000 AI images depicting national identities also highlighted these tendencies towards bias and stereotypes of TTI systems (e.g.~New Delhi's streets were mostly portrayed as polluted and littered) \cite{Turk2023}. This perpetuates cultural norms that are  prevalent in training datasets while under-representing less stereotypical and non-Western aspects of culture, art and society. Although some researchers have proposed ways to mitigate these effects, such as adding specific phrases (e.g. ``irrespective of gender'' \cite{bansal:emnlp2022}) or through the use of multilingual prompts (e.g. ``a photo of a king '' -- appending a Russian character to the prompt \cite{ventura:2023navigating}), these mitigation strategies are often ineffective (e.g. despite explicitly mentioning words such as ``white'', ``wealthy'' or ``mansion'', the authors in \cite{bianchi:facct23} report that Stable Diffusion continues to associate poverty with people of colour). In this paper we follow a different perspective from the aforementioned studies and perform the first large-scale analysis to understand the impact of prompting on art practice and art by analysing how prompting has changed over a 1 year timeframe.

\section{Datasets}
\label{s:datasets}

To perform our analysis we used three different datasets:
\begin{enumerate}
    \item \textbf{DiffusionDB}: a text-to-image prompt dataset containing 14 million images generated by Stable Diffusion, 1.8 million unique prompts, and hyperparameters \cite{wang-etal-2023-diffusiondb}. Collected over a 2 week period in August, 2022, the dataset is publicly available at: \url{https://poloclub.github.io/diffusiondb};
    \item \textbf{Midjourney 2022 Discord dataset}: 248k prompts and their associated generated images obtained by scraping ten public Discord channels over 28 days in June 2022 \cite{midjourneyDS-2022};
    \item \textbf{Midjourney 2023 dataset}: 2.84M prompts and associated generated images obtained by scraping public Discord channels over 16 days in October-November 2023. This dataset was created by the authors of this paper and is publicly available \cite{McCormack2024}.
\end{enumerate}

Due to the stochastic nature of the image generation process, it is common for prompting interfaces to produce a number of image variants from a single user supplied prompt. In Midjourney for example, the default is to generate four images, from which the user has options to regenerate, upscale, or produce four new variants from one of the generated images. For our prompt analysis we only consider the initial generation, removing identical prompts that appear in up-scaling or variant generation from our analysis. For the DiffusionDB dataset, tiled images were automatically split into separate images by the dataset's authors (one of the reasons why the number of images is much greater than the number of unique prompts).

Both the DiffusionDB and Midjourney 2022 Discord dataset (hereafter \textit{MJ\-2022}) were generated in the early stages of development and public release of these systems (around mid 2022). DiffusionDB was obtained by scraping public Discord channels when Stable Diffusion was in public beta testing. Midjourney has always used Discord as its interface for prompt generation.

Following an initial analysis of the DiffusionDB and MJ2022 datasets, we opted to create a new, more recent dataset to compare how prompting might have changed over the course of one year\footnote{Both Midjourney and Stable Diffusion have undergone several major developments since 2022. At the time of writing the current versions are Midjourney 5.2 and Stable Diffusion XL.}.

All the datasets include the full prompt text, user id, generated image URL and timestamp, the DiffusionDB dataset also includes other metadata, including NSFW scores for images and prompts, configuration parameters specific to stable diffusion and image dimensions. These additional fields were not used in our analysis.

\subsection{Initial Processing}
\label{ss:initial_processing}
Before analysing the datasets we performed some basic cleanups to help ensure the reliability of the analysis. This included removing any records with missing data or empty prompts, incomplete requests, and non-image generating prompts. Table \ref{t:basic_stats} gives a comparison of the three datasets used in our analysis. Prompts without validated users were removed from user statistics.

\begin{table}[]
    \centering
    \begin{tabular}{r | r r r}
     \textbf{Dataset}  &  \textbf{DiffusionDB} & \textbf{MJ2022} & \textbf{MJ2023} \\
     \hline \\
      Data collection period & 6-20 Oct 2022  & \hspace{1em}1-28 Jul 2022 & \hspace{1em}25 Oct-9 Nov 2023 \\
      Raw records            & 2.00M     & 250k    & 2.84M \\
      Prompts                & 1,528,513 & 145,080 & 936,589 \\
      Mean Prompt Length     & 162       & 150     &  101 \\
      \hline
      Users                  &  10,173   &  1,681   & 34,429\\
      Median prompts/user    & 54        & 12      & 8 \\
      Max prompts/user       & 5,630      & 2,493    & 3,666 \\
      \hline

    \end{tabular}
    \caption{Comparison of the three datasets used in this paper}
    \label{t:basic_stats}
\end{table}

\subsection{Dataset Statistics and Initial Analysis}
\label{ss:stats}

The most obvious change in the overall dataset comparisons shown in Table \ref{t:basic_stats} is the growth in the number of users -- Midjourney data from 2022 has less than 1.7k unique users, whereas just over a year later there are over 34k users, a twenty-fold increase. While there are more people using these TTI systems, they are not using the system as often or as much: the distribution of users vs. prompts shows a small number of ``power users'' (people spending a large majority of their time on the system), and a large number of casual users, with 75\% of users entering less than 15 prompts over the collection period in 2023.

Our analysis was performed using language tools based on English. However, we analysed the datasets to determine the language composition of prompts, using the fastText model \cite{joulin2016bag} %
to automatically classify each prompt by language. The results are shown in table \ref{t:languages}. As can be seen, English is by far the dominant language used in prompting within these systems, with over 99\% of prompts in each dataset identified as English\footnote{We classified non-text prompts -- such as Emojis -- as English. As the number of these type of prompts are very small, it did not impact significantly on the overall results.}. However, we note a significant increase in non-English prompts in languages such as French and Spanish between 2022 and 2023.

\begin{table}[]
    \centering
    \begin{tabular}{r | l r | l r | l r |}
     \multirow{2}{2em}{} &  \multicolumn{2}{|c|}{\textbf{DiffusionDB}} & \multicolumn{2}{|c|}{\textbf{Midjourney2022}} & \multicolumn{2}{|c|}{\textbf{Midjourney2023}} \\
     & Lang & Freq(\%) & Lang & Freq(\%) & Lang & Freq(\%) \\
     \hline \\
        1 & English & 99.914 & English & 99.893 & English & 99.13\\
        2 & French & 0.027 & French & 0.023 & French & 0.380\\
        3 & German & 0.012 & German & 0.016 & Spanish & 0.172\\
        4 & Spanish & 0.007 & Spanish & 0.013 & Italian & 0.091\\
        5 & Italian & 0.007 & Japanese & 0.012 & German & 0.072\\
        
       \hline
       \textbf{Total} & \multicolumn{2}{|r|}{50} & \multicolumn{2}{|r|}{19} & \multicolumn{2}{|r|}{14} \\
      \hline

    \end{tabular}
    \caption{Top five languages for prompts and the total number of languages detected in each dataset}
    \label{t:languages}
\end{table}

Prompts were divided into components (\emph{specifiers}), separated by commas or periods, ignoring letter case and removing a small set of stopwords, system commands and punctuation (similar to \cite{sanchez:2023}). We also ensured that terms utilising periods (e.g.~f1.2) are captured. Table \ref{t:top_10} shows the most popular individual specifiers for each dataset and the total number of unique specifiers. Frequency values are given as a percentage of the total number of specifiers in each dataset's corpus. For the Midjourney datasets, we differentiate between \emph{initial} and \emph{variation} prompts (typically initiated using the \texttt{/imagine} command) and other requests such as up-scaling or panning/zooming. For the prompt analysis we only consider the initial and variation prompts.

\begin{table}[htbp]
    \centering
    \begin{tabular}{r | l r | l r | l r |}
    \toprule
     \multirow{2}{4em}{} &  \multicolumn{2}{|c|}{\textbf{DiffusionDB}} & \multicolumn{2}{|c|}{\textbf{MJ2022}} & \multicolumn{2}{|c|}{\textbf{MJ2023}} \\
     & Term & Freq(\%) & Term & Freq(\%) & Term & Freq(\%)\\
     \hline
   1 & highly detailed        & 11.19 & cinematic               & 7.21 & white background & 2.76 \\
   2 & artstation             & 10.00 & octane render           & 5.98 & 8k & 2.72 \\
   3 & sharp focus            &  8.80 & 8k                      & 5.43 & cinematic & 2.11 \\
   4 & trending on artstation &  8.39 & artstation              & 4.66 & photorealistic & 1.99 \\
   5 & concept art            &  8.34 & ar 16:9                 & 4.40 & realistic & 1.82 \\
   6 & octane render          &  6.56 & 4k                      & 3.95 & 4k & 1.75 \\
   7 & intricate              &  6.54 & detailed                & 3.30 & black and white & 1.35 \\
   8 & digital painting       &  6.41 & trending on artstation  & 2.51 & hyper realistic & 1.24 \\
   9 & illustration           &  6.06 & realistic               & 2.29 & raw & 1.21 \\
  10 & 8 k                    &  6.02 & highly detailed         & 2.17 & cinematic lighting & 1.12 \\
       \hline
       \textbf{Total} & \multicolumn{2}{|r|}{1,723,539} & \multicolumn{2}{|r|}{156,045} & \multicolumn{2}{|r|}{512,634} \\
      \hline
    \end{tabular}
    \caption{The 10 most frequent specifiers and total number of unique specifiers for each dataset}
    \label{t:top_10}
\end{table}

Phrases to cajole the image quality or specific style dominate, reflecting common folklore in prompting that phrases such as ``cinematic'', ``highly detailed'' and ``8k'' result in better images than those without such qualifiers. Interestingly, the terms ``artstation''\footnote{Artstation is a portfolio showcase site, popular for games and commercial entertainment artists.} and ``trending on artstation'' feature prominently in the 2022 datasets, but are no longer in the top 10 by 2023. The likely reason for this is that early versions of Stable Diffusion suggested using these (and other terms in the top 10) in their introductory prompt guides. In early versions of Stable Diffusion in particular, these terms were necessary to get images of good quality, but are no longer required in the most recent releases.

\begin{table}[]
    \centering
\begin{tabularx}{\textwidth} {>{\raggedright\arraybackslash}X >{\small}X >{\small}X }

     \textbf{DiffusionDB} & \textbf{MJ2022} & \textbf{MJ2023}  \\
\hline
    style of beksinski  & machinarium       & cartoon \\
    anime               & pixar             & anime \\
    fantasy             & manga             & pixar \\
    pixar               & in stylized style & watercolor \\
    cyberpunk           & anime             &  comic (book) \\
    vogue cover         & anime waifu style character & vector \\
    smooth style beeple & photographic      & banana fish anime \\
\hline
\end{tabularx}
    \caption{Popular prompt specifiers related to style or stylistic direction}
    \label{t:styles}
\end{table}

We analysed the datasets for references to styles. A selection of the most popular styles for each dataset are shown in Table \ref{t:styles}.
As can be seen from the table, illustration styles (``anime'') or names of popular studios (``Pixar'') are the most common. We observe a shift from the general (``anime'', ``fantasy'') to the specific (``anime waifu style character'', ``banana fish anime''), suggesting that users are after more specific stylistic references in the images produced. While not shown in the table, cultural memes of the time also feature strongly, with references to ``Barbie'' featuring in the MJ2023 dataset, for example. The final observation is the diversity of styles increases with the number of users of the system, with more generic styles dominating the MJ2023 dataset (``cartoon'', ``watercolor'', ``comic book'').

\section{Analysis}
\label{s:analysis}

\subsection{Use of Artist Names}
\label{ss:artist_names}

\begin{quote}
``Stability has a music generator that only uses royalty free music in their dataset. Their words: ``Because diffusion models are prone to memorisation and overfitting, releasing a model trained on copyrighted data could potentially result in legal issues.'' Why is the work of visual artists being treated differently?''\cite{loisBlog} (Lois Van Baarle -- \verb|#|3 artist in DiffusionDB).     
\end{quote}

It has become a common practice to use artist's names in prompts as a way of generating images in the style of that artist -- what has been called ``style theft'' \cite{McCormack:2023aa}. We used the SpaCy\footnote{\url{https://spacy.io}} NLP library to perform identification of people's names in the datasets. We also looked at specifiers that contain the term ``style'', as often desired styles reference artworks or organisations rather than creators (e.g. ``Banana Fish anime style'' or ``Pixar style'' -- Table \ref{t:styles}).

Tables \ref{t:topArtistsSDDB}, \ref{tab:topArtistsMJ22} and \ref{tab:topArtistsMJ23} show demographic information for the top ten artists mentioned in prompts, for DifussionDB, MJ2022 and MJ2023 datasets respectively. In all datasets, male artists dominate requests for specific artists' styles. Similarly, artists from the United States form the largest group, and artists from Western cultures dominating in general.

In the earlier datasets (DiffusionDB and MJ2022), fantasy, comic, and game art styles dominates across the top artists. This trend reflects the demographics, aims and use-cases of early adopters of the system. In the MJ2023 dataset, we see that the user base has grown significantly. Accordingly, the style of top artists has moved more into the realm of mainstream art, seen by the increase in the range of art styles including film, abstract painting, photography, sculpture, and architecture (as evidenced by Tables \ref{t:topArtistsSDDB}, \ref{tab:topArtistsMJ22} and \ref{tab:topArtistsMJ23}) -- with realism emerging as the prevalent artistic style. We also see a shift towards a greater balance between contemporary and modern artists in the later dataset (as evidenced by the active years of artists in the latter dataset). We believe this difference follows the more widespread adoption of the tool over time.

\begin{table}[ht]
\centering
\begin{tabular}{@{}llllll@{}}
\toprule
     \textbf{Artist Name} & \textbf{Freq(\%)} & \textbf{Gender} & \textbf{Based in} & \textbf{Age} & \textbf{Medium} \\
\midrule 
      \href{https://rutkowski.artstation.com/}{Greg Rutkowski} & 9.58 & M & POL & 35 & Illustrator \& Concept Artist \\
      
      \href{https://www.jamesjean.com/}{James Jean} & 1.43 & M & USA & 44 & Illustrator \& Painter \\
      
      \href{https://loish.net/about/#}{Lois Van Baarle} & 1.12 & F & NLD & 38 & Digital Artist \& Illustrator \\
      
      \href{https://www.angelarium.net/}{Peter Mohrbacher} & 1.10 & M & USA & 40 & Concept artist \& Illustrator \\ 

      \href{http://www.mostlywanted.com/}{Tom Bagshaw} & 0.86 & M & GBR & 46 & Illustrator \\

      \href{https://www.goodbrush.com/}{Craig Mullins} & 0.80 & M & USA & 59 & Concept artist \& Digital Painter \\

       \href{https://rossdraws.com/}{Ross Tran} & 0.75 & M & USA & 31 & Concept Designer \& Illustrator \\

       \href{https://www.ruanjia.com/}{Ruan Jia} & 0.75 & M & CHN & 40 & Concept \& Digital artist \\

       \href{https://www.dan-mumford.com/}{Dan Mumford} & 0.74 & M & UK & 32 & Illustrator \& Concept Artist\\
      
       \href{https://en.wikipedia.org/wiki/Norman_Rockwell}{Norman Rockwell}\text{*} & 0.68 & M & USA & 84 & Painter \& Illustrator \\
     
\bottomrule
    \end{tabular}
    \caption{Top 10 artists named in prompts from the DiffusionDB dataset, the \textbf{Freq} column shows the frequency, in percent, that the name appears in all prompts that mention a name. Names with a \text{*} symbol denote artists that are deceased.}
    \label{t:topArtistsSDDB}
\end{table}

\begin{table}[ht]
\centering
\begin{tabular}{@{}llllll@{}}
\toprule
 \textbf{Artist Name} & \textbf{Freq(\%)} & \textbf{Gender} & \textbf{Based in} & \textbf{Age} & \textbf{Medium} \\ 
 \midrule

\href{https://www.goodbrush.com/}{Craig Mullens} & 4.10   & M   & USA      & 59 & Digital Painter \& Illustrator \\
\href{https://www.angelarium.net}{Peter Mohrbacher} & 2.73  & M   & USA      & 40  & Digital Painter \& Illustrator \\
\href{https://en.wikipedia.org/wiki/Tsutomu_Nihei}{Tsutomu Nihei}  & 1.61   & M   & JPN    & 52  & Manga Artist                 \\
\href{https://en.wikipedia.org/wiki/Robert_Mapplethorpe}{Robert Mapplethorpe}\text{*} & 1.42 & M   & USA      & 43 & Photographer       \\
\href{https://www.jamesjean.com}{James Jean}  & 1.22      & M   & USA      & 44 & Painter \& Illustrator       \\
\href{https://en.wikipedia.org/wiki/Albert_Bierstadt}{Albert Bierstadt}\text{*} & 0.94 & M & GER/USA & 72 & Painter \\
\href{https://en.wikipedia.org/wiki/Katsuhiro_Otomo}{Katsuhiro Otomo} & 0.91  & M   & JPN        & 69 & Manga Artist \& Illustrator \\ 
\href{https://art.marcsimonetti.com}{Marc Simonetti} & 0.91   & M   & FRA        & 46 & Conceptual Artist \& Illustrator \\ 
\href{https://www.rossdraws.com}{Ross Tran}   & 0.89      & M   & USA      & 31  & Concept Artist \& Illustrator  \\
\href{http://www.artofmikemignola.com}{Mike Mignola}  & 0.80    & M   & USA      & 63 & Comic Artist \& Writer        \\ 
\bottomrule
\end{tabular}
\caption{Table of the top 10 artists mentioned in the MidJourney 2022 dataset (names with a \text{*} symbol denote artists that are deceased).}
\label{tab:topArtistsMJ22}
\end{table}

\begin{table}[ht]
\centering
\begin{tabular}{@{}llllll@{}}
\toprule
 \textbf{Artist Name} & \textbf{Freq(\%)} & \textbf{Gender} & \textbf{Based in} & \textbf{Age} & \textbf{Medium} \\
\midrule
    \href{https://en.wikipedia.org/wiki/Wes_Anderson}{Wes Anderson} & 0.51 & M & USA & 54 & Film director \& Producer \\
    \href{http://frankfrazetta.net/index.html}{Frank Frazetta}\text{*} & 0.50 & M & USA & 82 & Illustrator \\
    \href{https://en.wikipedia.org/wiki/Gertrude_Abercrombie}{Gertrude Abercrombie}\text{*} & 0.28 & F & USA & 68 & Painter \\
    \href{https://en.wikipedia.org/wiki/William_Baziotes}{William Baziotes}\text{*} & 0.27 & M & USA & 50 & Painter \\   
    \href{https://en.wikipedia.org/wiki/David_Smith_(sculptor)}{David Smith}\text{*} & 0.27 & M & US & 59 & Sculptor \& Painter \\      
    \href{https://en.wikipedia.org/wiki/Yayoi_Kusama}{Yayoi Kusama} & 0.25 & F & JPN & 94 & Sculptor, Art Installation \& Painter \\
    \href{https://www.timburton.com/}{Tim Burton} & 0.24 & M & USA & 65 & Film Maker \& Animator \\
    \href{https://en.wikipedia.org/wiki/Alphonse_Mucha}{Alphonse Mucha}\text{*} & 0.24 & M & CZE & 78 & Painter, Illustrator \& Graphic Artist \\
    \href{https://www.maxdunbar.com/}{Max Dunbar} & 0.23 & M & CAN & -- & Illustrator, Comic \& Concept Artist \\
    \href{https://www.instagram.com/studioanneholtrop/}{Anne Holtrop} & 0.23 & M & NLD & 46 & Architect \\
\bottomrule
\end{tabular}
\caption{Table of the top 10 artists mentioned in the MidJourney 2023 dataset (names with a \text{*} symbol denote artists that are deceased).}
\label{tab:topArtistsMJ23}
\end{table}


As images `in the style of' specific artists become prolific, some artists are pursuing legal action or voicing their concerns and disapproval about the non-consensual use of their work in training generative AI systems\footnote{In late 2022 artists started a public mass protest against AI-generated artwork on ArtStation \url{https://arstechnica.com/information-technology/2022/12/artstation-artists-stage-mass-protest-against-ai-generated-artwork/}}. A major issue -- currently disputed -- is that of copyright \cite{chesterman:2023}. An argument put forward is that industries such as music have long-established rights in place to prevent on-line copying and protect copyright. In response, companies such as Stability AI claim that generative AI is not about copying artworks but creating new ones \cite{lemley:2023,guadamuz:2023}, as is the common practice of human artists who seek inspiration by viewing or copying other artists' work. Whether or not this dispute rules in favour of visual artists, it may involve more than just removing their original work from the training data, it may also require to remove all images that have been influenced by this artist's style as TTI systems work from the collective use of data \cite{intellectualProperty:2023}. 

Artists whose styles are being copied by AI also face criticism due to the perceived ease by which AI can replicate their style, with critics arguing that this ``exposes weaknesses’' of human-made art \cite{criticArticle} and emphasising the possibility that AI art may lead individuals to be ``more rigorous enjoyers/audiences of art'' \cite{peterMohrbacherArticle}. Although art critique is a standard practice in the art world, this fails to acknowledge what is beyond the images themselves. As Wes Anderson (\verb|#|1 artist in MJ2023) puts it, ``An artist or illustrator has a particular hand that they've developed and they find their set of ideas -- they find their voice\ldots And I don’t know how good AI is at creating a voice’’ \cite{andersonArticle}. Some argue that an increased engagement of the artist with the audience would be needed in order to mitigate these issues \cite{lyu:2022}.

Contrary to the above views, some of the artists whose work has been heavily featured in training datasets have identified potential roles for TTI systems in their creative practice. Peter Mohrbacher (\verb|#|2 artist in MJ2022) mentioned using Midjourney as a kind of ``personalised'' image generation model \cite{peterMohrbacherArticle}, while 
artist Ruan Jia (\verb|#|8 artist in DifussionDB) suggested the use of TTI systems as a communication and brainstorming tool within artistic teams \cite{ruanArticle}.
When thinking about the artistic possibilities of TTI systems, emphasising the dialogue between human artists and machines rather than just the final artefact may be more insightful \cite{oppenlaender:2022}; in this sense, creating a unique style would require more than just entering a prompt. 

Looking beyond the individual, one may ask how stylistic referencing might impact art and culture more generally. The more people are exposed to certain bodies of work, the more these works influence their preferences: ``acquired cultural competences play a role in art preference; the more familiar the art and artist are to the consumer and the more s/he knows about them, the higher satisfaction the art gives.'' \cite{uusitalo:2009}. The ease with which one can generate content using these systems, and the ease of faithfully replicating a visual style, may play a significant role in influencing the prominence of certain artists or artistic styles.

\subsection{Topic Analysis}
\begin{figure}
    \centering
    \includegraphics[width=\textwidth]{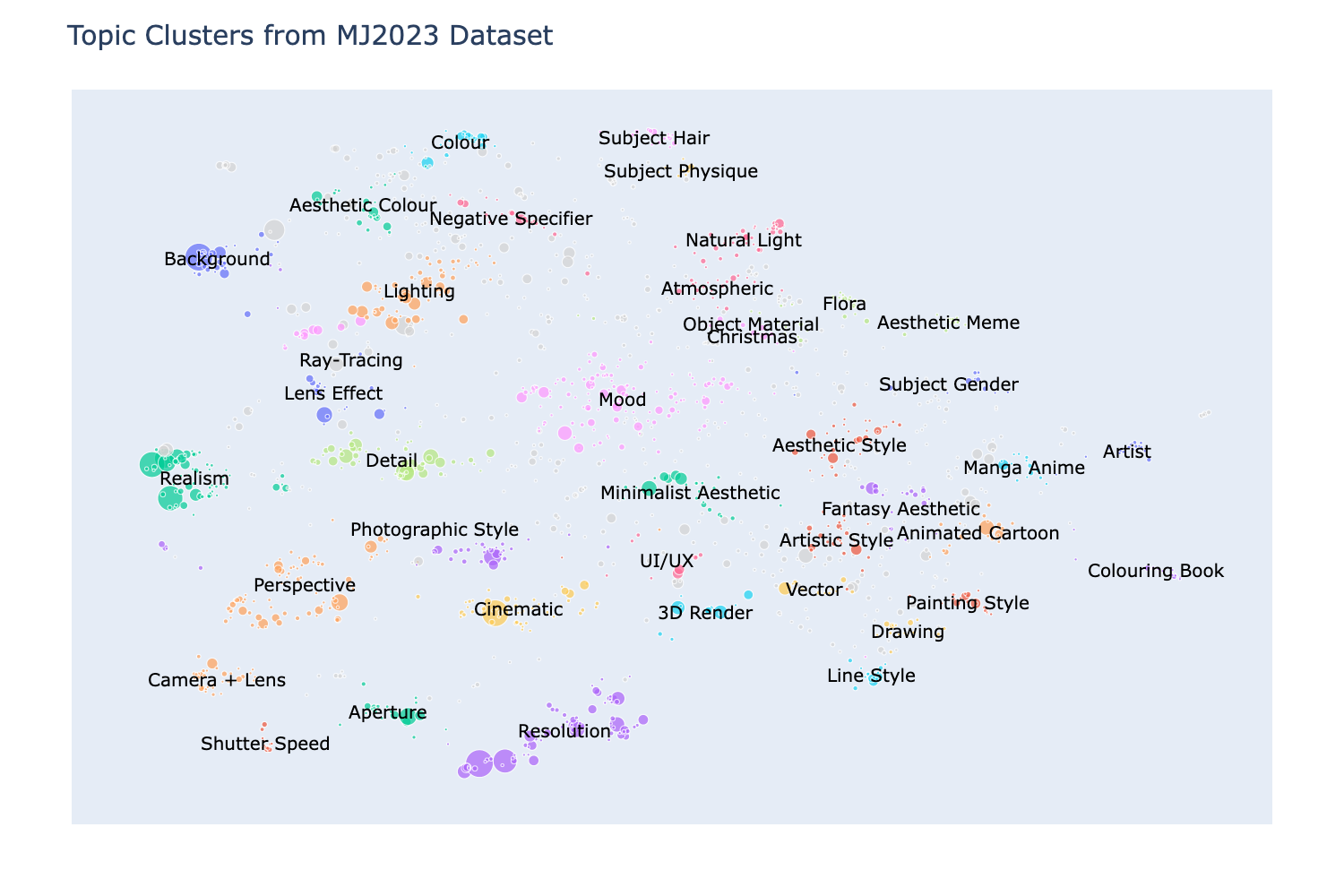}
    \caption{Visualisation of the MPnet embeddings of 1700 prompt specifiers and the 40 topics identified using the HDBSCAN clustering algorithm.}
    \label{fig:topic-analysis}
\end{figure}

Building on a previous topic analysis of the DiffusionDB dataset \cite{sanchez:2023}, we conducted a topic analysis on prompt specifiers in the MJ2023 dataset.  Prompt specifiers with 100 or more uses were included in the analysis, resulting in 1700 individual prompt specifiers. The topic modelling adhered to the approach outlined in \cite{grootendorst2022bertopic} and involved first utilising the MPnet (Masked and Permuted Pre-training for Language Understanding) model \cite{MPnet} to encode prompt specifiers into vector representations. MPnet, a language model specialising in language understanding, is capable of embedding phrases into a latent space that proves effective for diverse language tasks, including clustering and sentence similarity. As in \cite{sanchez:2023}, we utilised the \textit{all-mpnet-base-v2} pre-trained weights and the sentence-transformer library \cite{sbertdocs} for encoding. The UMAP \cite{mcinnes2020umap} dimensionality algorithm was then applied to the vector embeddings to reduce them to a 5-dimensional space in order to mitigate the negative effect of high dimensionality on clustering performance \cite{VandenBusscheJan2000OtSB}. We utilised, the hierarchical density-based spatial clustering (HDBSCAN) algorithm \cite{campello2013density} to identify topic clusters within the embedding space resulting in an initial collection of 40 topics. We present these topics in figure \ref{fig:topic-analysis} where we utilised UMAP to reduce the original embedding to 2D. It is worth noting that there are some limitations to this method, with some outliers being assigned to incorrect clusters. However, as in \cite{sanchez:2023}, we utilised a class-specific term frequency-inverse document frequency (c-TF-IDF) to identify the most important specifiers in each group and utilised them to accurately label the cluster. An interactive version of Figure \ref{fig:topic-analysis} can be found at \footnote{\url{https://sensilab.github.io/NoLongerTrendingOnArtstation/}}.

The five most referenced topics were \textit{Resolution}, \textit{Realism}, \textit{Mood}, \textit{Detail} and \textit{Lighting}, which account for roughly a third of prompt specifiers. This is expected, given the fundamental nature of these topics in the context of images. Many topics found in this analysis were also present in \cite{sanchez:2023}, such as the fantasy aesthetic and photography themes. 

A prominent theme identified in our analysis was photography, encompassing topics such as \textit{Camera + Lens}, \textit{Shutter Speed}, \textit{Aperture}, \textit{Photographic Style}, \textit{Perspective}. This, combined with the widespread use of \textit{Realism} specifiers, indicates that users are utilising MidJourney to generate images that are photo realistic. Moreover, technical camera functions, such as those related to lens aperture and camera shutter speed, are being applied as if they were operating on a physical camera. This may be an attempt to control visual features such as depth of field and motion blur in both photorealistic and non-realistic images.

Art was another identified theme and covered topics such as \textit{Painting Style}, \textit{Drawing}, \textit{Artistic Style} ect. This suggests that the system is still being used by some as a tool to explore aesthetics or to generate images that mimic a particular artistic style or medium.

Some unexpected topics discovered in our analysis included \textit{Christmas} and \textit{Colouring Book}. We hypothesise that \textit{Christmas} emerged as a topic due to the data being collected in proximity to Christmas, perhaps inspiring users to generate Christmas themed images. \textit{Colouring Book} illustrates how users are prompting the system to generate assets that have value beyond their static form and can be used in more recreational settings, such as colouring for children and adults.

\section{Coda: What are all these images actually of?}
\label{s:coda}

So far we have looked at prompting and its implications through the lens of the prompts themselves, where language is used as a proxy for the user's intent. A prompt writer must craft the prompt to get the image they want and the success (or otherwise) of that prompt is evaluated \emph{visually} rather than linguistically. As we have discussed, prompting is a rather convoluted process, requiring specific keywords around style and surface aesthetics, which tend to dominate our analysis of over 3M prompts from the three datasets.

\begin{table}[]
    \centering
    \tiny
    \begin{tabularx}{\textwidth} { 
  | >{\raggedright\arraybackslash}X 
   >{\centering\arraybackslash}X 
   >{\raggedright\arraybackslash}X | }
  \hline
    \small\textbf{Prompt}    & \small\textbf{Generated Image} & \small\textbf{Description}  \\
    \hline
     \texttt{Imagine a dream-like scene where reality blurs and the boundaries between woman and peacock dissolve. Sketch a woman's body full of delicate vulnerability, her features soft and poetic. Let the peacock's head emerge, seamlessly integrating with its essence, symbolizing the deep connection with the world of colors of the peacock's tail. Use the empasto technique to add a tactile quality, allowing the viewer to visually feel the texture of the artwork. Set against a deep, velvety canvas of dark blue on a black background, this ethereal combination creates a sense of enchantment, encouraging viewers to explore the depths of their imagination}    & \raisebox{-1.0\totalheight}{\includegraphics[width=0.3\textwidth]{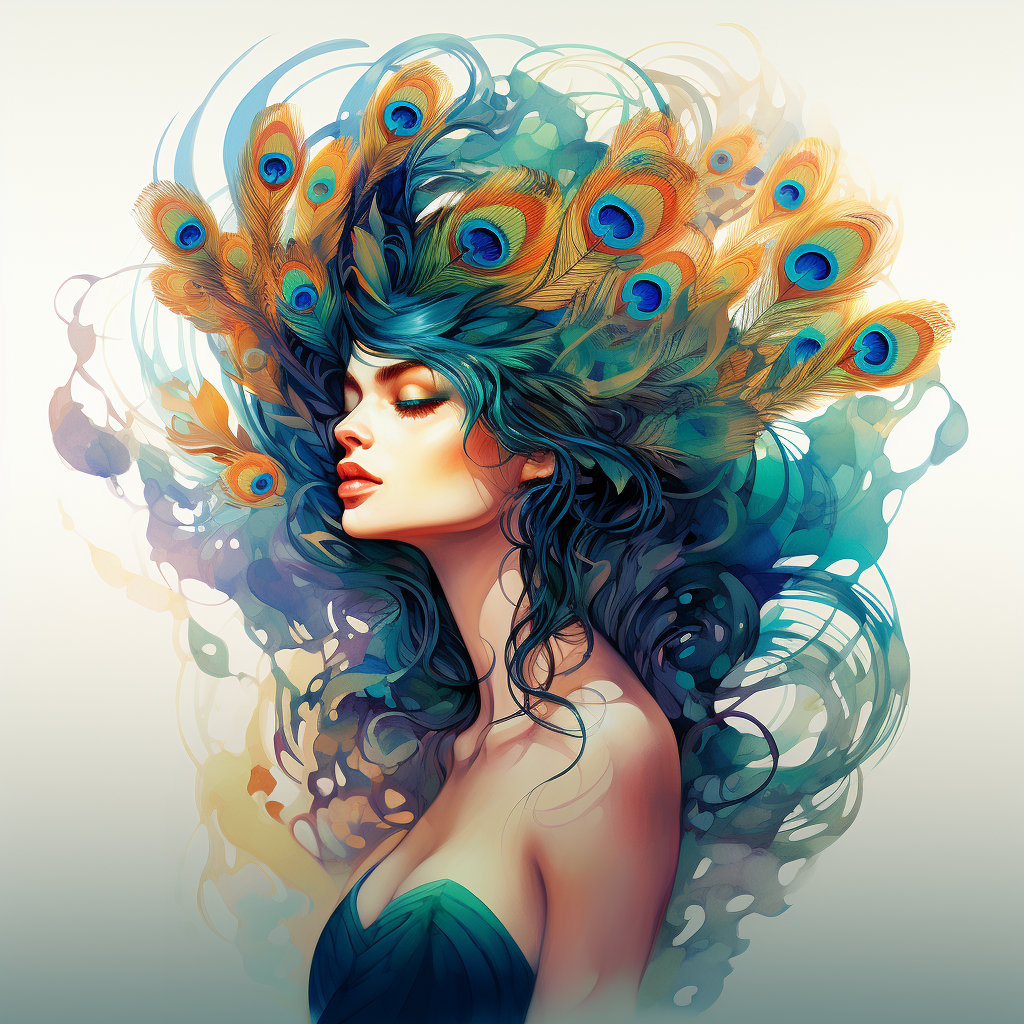}} & \textit{painting of a woman with peacock feathers on her head}\\
     \hline
  \end{tabularx}
    \caption{Descriptive captioning example of a prompt from the MJ2023 dataset. The prompt (left) was used to generated the image shown (middle), with resultant description (right) provided by the BLIP model.}
    \label{t:blip_example}
\end{table}

While language analysis tells us something about the prompt-author's intent, it does not tell us if that intent was successfully realised in the resultant image. To better understand this aspect, we turned our attention to ``upscale'' requests in the dataset. The reasoning is that you will only bother to upscale an image if it is of interest, i.e.~one or more of the image variants from the prompt you specified may be approaching what you intended (or like). This is a typical workflow for Midjourney (recall that our prompt analysis was performed on initial or variation requests, not on upscale requests).

We used the upscale requests from the MJ2023 dataset. After removing duplicates and failed requests, a total of 536,014 prompts and associated AI generated images were found. To discover what images actually depict, we ran each image through a state-of-the-art BLIP (Bootstrapping Language-Image Pre-training for Unified Vision-Language Understanding and Generation) image captioning model \cite{li2022blip}, generating descriptive captions for each image. The captioning model gives an overall description of the image, much like the way a human would, as the model recognises not only the objects in a scene but the relationships and basic surface aesthetic properties. We limited the description to a maximum of 20 words to avoid overly wordy descriptions with unnecessary detail. An example is shown in Table \ref{t:blip_example}.

We then analysed each description to find the most popular depictions in the generated images. After removing adjectives and basic aesthetic or stylistic words (e.g.~references to colours) the top subjects depicted were: \textit{woman} (22.26\% of images), \textit{man} (16.2\%), \textit{dress} (6.92)\%, \textit{hair} (5.51)\%
\textit{room} (5.44)\%
\textit{flower} (5.33)\%.

\begin{figure}
    \centering
    \includegraphics[width=0.9\textwidth]{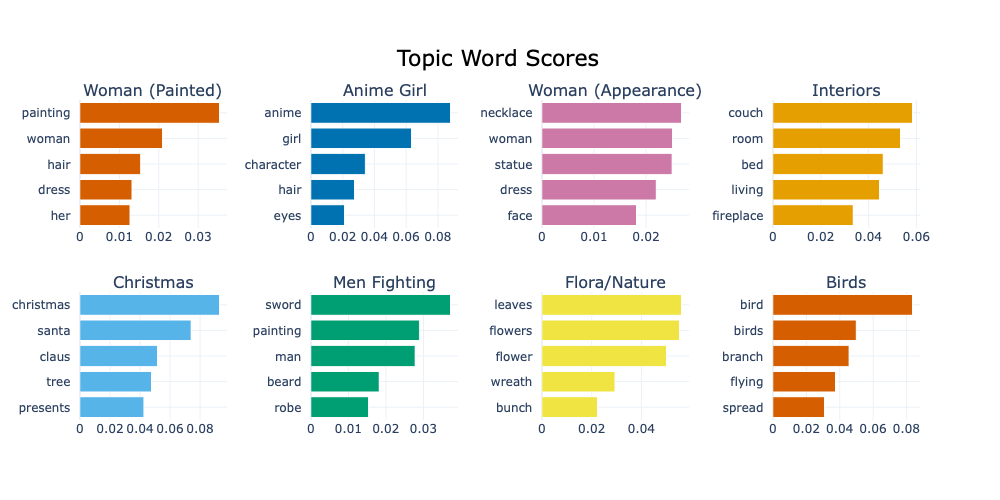}
    \caption{The 8 most popular topics for upscaled images in MJ2023}
    \label{f:image_topics}
\end{figure}

We next performed topic modelling on the description data using BERTopic \cite{sbertdocs}. Similar to the prompt Topic Analysis, image descriptions were converted to embeddings then dimensionally reduced using UMAP before clustering with HDBSCAN. BERTopic uses cTF-IDF to get topic weightings before an additional fine-tuning to optimise the number of topic categories. Figure \ref{f:image_topics} shows the eight most important topics and relative frequency of keywords.

Our analysis shows that the most dominant image subject for MidJourney users are illustrations of Women and Anime Girls. 

\section{Summary and Conclusion}
\label{sec:discussion}

\subsection{Limitations of the study}
\label{ss:limitations}

Our datasets represent results only for two systems: Stable Diffusion and Midjourney. Other popular systems, such as DALL-E, Leonardo, Crayion, Firefly are all propriety, making access to prompts difficult. As Stable Diffusion is an open-source model, it is no longer run via Discord and users are free to download their own versions. One possibility to address this issue would be to scrape data from the many ``prompt showcase'' websites that communities of people can post AI-generated images and their associated prompts (e.g.~sites such as {Prompt Hero} \url{https://prompthero.com}), we leave this as future work.

\subsection{Findings}
\label{ss:findings}

As shown in Table \ref{t:basic_stats}, Midjourney's user base has grown significantly between 2022 and 2023. The median prompt length is reducing, probably due to changes in the software that require less specialist prompting (hinting at the tool's accessibility). While there is a significant increase in users, the majority of those users do not prompt very much, suggesting that for many people, the use of TTI systems is largely recreational (something confirmed informally by a poll in \cite{Kelly2022}). This observation is further reinforced from our topic analysis of both prompts and images, with topics such as ``colouring books'' and ``Christmas'' being popular (MJ2023 data was collected in November, leading to the Christmas period).

Our analysis shows that prompting emphasises and privileges popular styles and surface aesthetic appearances (``cinematic lighting'', ``photorealistic'', ``ultra detailed''). It forces a fixation on surface aesthetics, potentially at the expense of other important factors in art making, including narrative, realism, authenticity and individuality. 
    
While we identified a wide variety of artistic styles, TTI systems tend towards the popular, reinforcing stylistic norms and aesthetic ``sameness''. Our image analysis showed what seems like common knowledge when viewing websites such as the Midjourney Showcase or various ``Prompt Art'' sites: that most of the images are closeups or medium shots of young women. Genres of fantasy art, game art and comic or anime illustration dominate the specifiers used in prompting. As prior research has demonstrated, TTI systems continue to promote and reinforce racial, gender and other cultural biases. Despite the vast volume of new synthetic images generative AI has brought us, its seems that most people using these systems are after ``more of the same''. In the process of working with these datasets and being exposed to many thousands of images, our view is that few if any AI images are memorable in the way that human art is, and its difficult to see how they can be aesthetically unique in the way that human art can.

The agency exerted by TTI systems onto the human user, while difficult to quantify, is an important topic for future investigation. When viewing TTI systems as a creative medium, their inherent properties will inevitably shape and contribute to the future of image production. The precise way in which this occurs is multiple and nuanced.  In this paper, we set out to trace this effect through a qualitative interpretation of a statistical analysis of the language employed in TTI systems and the images they produce. 
This leads us to conclude that generative AI imagery, at least in its current form, is probably not a serious threat to human art. After all, surface mimicry of a popular style does not constitute an artistic innovation or practice. A physical painting will always embody the act and intention of a human artist, something that no synthetic AI image can ever do.

\subsubsection{Acknowledgements}
This research was supported by an Australian Research Council Grant, DP220101223.

%
%
%
\bibliographystyle{splncs04}
\bibliography{McCormack}
\end{document}